\documentclass[aps,prl,twocolumn, superscriptaddress,groupedaddress]{revtex4}               

\usepackage{graphicx}
\usepackage{amsmath}
\usepackage{amssymb}
\usepackage[amssymb]{SIunits}
\usepackage{color}
\usepackage[usenames,dvipsnames]{xcolor}
\usepackage{ulem}
\usepackage{cancel}
\usepackage{natbib}
\usepackage[T1]{fontenc}
\usepackage{bm}
\usepackage{graphicx}
\usepackage{epstopdf, epsfig}
\usepackage{color}
\usepackage{soul}
\usepackage{subfigure}
\usepackage{amsmath}
\usepackage{lineno}
\usepackage{setspace}
\usepackage{longtable}
\makeatletter

\begin{document}
\setstretch{1.0} 
\title{Evidences for the existence of the ultimate regime in supergravitational turbulent thermal convection}

\author{Hechuan Jiang}
\thanks{Equally contributed authors}
\affiliation{Center for Combustion Energy, Key Laboratory for Thermal Science and Power Engineering of MoE, and Department of Energy and Power Engineering, Tsinghua University, 100084 Beijing, China.}

\author{Dongpu Wang}
\thanks{Equally contributed authors}
\affiliation{Center for Combustion Energy, Key Laboratory for Thermal Science and Power Engineering of MoE, and Department of Energy and Power Engineering, Tsinghua University, 100084 Beijing, China.}

\author{Shuang Liu}
\affiliation{Center for Combustion Energy, Key Laboratory for Thermal Science and Power Engineering of MoE, and Department of Energy and Power Engineering, Tsinghua University, 100084 Beijing, China.}

\author{Chao Sun}
\thanks{chaosun@tsinghua.edu.cn}
\affiliation{Center for Combustion Energy, Key Laboratory for Thermal Science and Power Engineering of MoE, and Department of Energy and Power Engineering, Tsinghua University, 100084 Beijing, China.}
\affiliation{Department of Engineering Mechanics, School of Aerospace Engineering, Tsinghua University, 100084 Beijing, China.}

\date{\today}

\begin{abstract} 
{ What is the final state of turbulence when the driving parameter approaches to infinity? 
	For thermal turbulence, in 1962, Kraichnan proposed a so-called ultimate scaling dependence of the heat transport (quantified by the Nusselt number $\text{Nu}$) on the Rayleigh number ($\text{Ra}$), which can be extrapolated to arbitrarily high $\text{Ra}$. The existence of Kraichnan ultimate scaling has been intensively debated in the past decades. In this work, using a supergravitational thermal convection system, with an effective gravity up to 100 times the Earth's gravity, both Rayleigh number and shear Reynolds number can be boosted due to the increase of the buoyancy driving and the additional Coriolis forces. 
	Over a decade of $\text{Ra}$ range, we demonstrate the existence of Kraichnan-like ultimate regime with four direct evidences: the ultimate scaling dependence of $\text{Nu}$ versus $\text{Ra}$; the appearance of turbulent velocity boundary layer profile; the enhanced strength of the shear Reynolds number; the new statistical properties of local temperature fluctuations. 
		The present findings will greatly improve the understanding of the flow dynamics in geophysical and astrophysical flows.  }
\end{abstract}

\maketitle
\section{Introduction}
Thermally driven turbulent flows ubiquitously occur in meteorological \cite{atmosphere1}, geophysical \cite{Intro_geophy_mantle1,Intro_geophy_mantle2} and astrophysical \cite{Intro_Sun_1,Intro_Sun_2} flows and large scale industrial processes \cite{engineering}. A paradigm for modeling the thermally driven turbulent flows is Rayleigh-B\'{e}nard convection (RBC) (see \cite{Review1,Review2,Review3,Review4} for reviews), which is a layer of fluid confined between two horizontal plates heated from below and cooled from above. { The direction of gravity is the same as the temperature gradient.} In those aforementioned natural phenomena, the driving strength of the flow is extremely high and far beyond the accessible regime in the lab scales. A powerful approach for the investigation of {these natural phenomena and industrial processes} using laboratory experiments is to find the asymptotic laws that can enable us to extrapolate the laboratory findings to the unattainable parameter regimes in the natural {and industrial} systems. One example of such asymptotic laws is on the dependence of the Nusselt number $\text{Nu}$  (dimensionless convective heat flux) on the Rayleigh number $\text{Ra}$ (dimensionless strength of driving buoyancy) in the limit of intense thermal forcing. 
In 1962, Kraichnan predicted as $\text{Ra}$ increases, thermal convection will reach an  ultimate state with $\text{Nu} \propto \text{Ra}^{1/2}{(\ln \text{Ra})^{ - 3/2}} \propto \text{Ra}^ {\gamma}$  \cite{Kraichnan1962}, which yields a steeper effective scaling exponent ($\gamma >$ 1/3, and asymptotically approaches to 1/2) as compared to the scaling exponent ($\gamma \lesssim$ 1/3) in the classical state {before the transition, where the turbulent transport is limited by laminar boundary layers (BLs)}.

In the past 40 years, tremendous amount of efforts have been put into to search for this ultimate scaling in high $\text{Ra}$ regimes \cite{Castaing1989,Chavanne1997,Niemela2000,niemela2010,Roche2010,He2012,Urban2014,Ahlers2017,Sreenivasan2020pnas,he2020pnas,kawano2021}. For the reported experimental results, most $\text{Nu}(\text{Ra})$ measurements are consistent below $\text{Ra} \simeq {10^{11}-10^{12}}$, while the situation at larger $\text{Ra}$ becomes puzzling. {At $\rm{Ra} \gtrsim 10^{11}$, the compensated heat transport $\rm{Nu} \cdot \rm{R{a^{ - 1/3}}}$ decreases or levels out with Ra in some studies \cite{Castaing1989,Niemela2000,niemela2010,Urban2014,Sreenivasan2020pnas} while it increases  in others \cite{Chavanne1997,Roche2010,He2012,Ahlers2017,he2020pnas,kawano2021}.}
For numerical simulations, 
limited by the power of supercomputers, three dimensional direct numerical simulations (DNS) can only reach $\text{Ra} = 2 \times {10^{12}}$ for a moderate aspect ratio {between the diameter and height of the convection cell} \cite{stevens2011}, which does not show the transition to the ultimate regime.
For both experiments and simulations, it is notoriously difficult to arrive at very high Ra while keeping other parameters to be constants.
{Recently, signatures of the transition to ultimate regime have been reported in the experimental studies of traditional RBC using a large cylindrical sample \cite{He2012}.}

In this work, we explore the ultimate regime of thermal turbulence in a supergravitational thermal convection system through rapid rotation. {An annular cylinder subject to a radial temperature gradient has been used to study baroclinic waves, zonal flow and convection in the low Ra regime \cite{Fowlis1965,busse_1974,Hide1975,busse_1985,busse_1986a,busse_1986b,busse_1994,Read2017,Kang2017,Kang2019,Menaut2019}. Here, we exploit the rapidly rotating system to study high Ra thermal turbulence.}

\section{Experiments and Simulations}

\begin{figure*}
	\centerline{\includegraphics[width=0.99\linewidth]{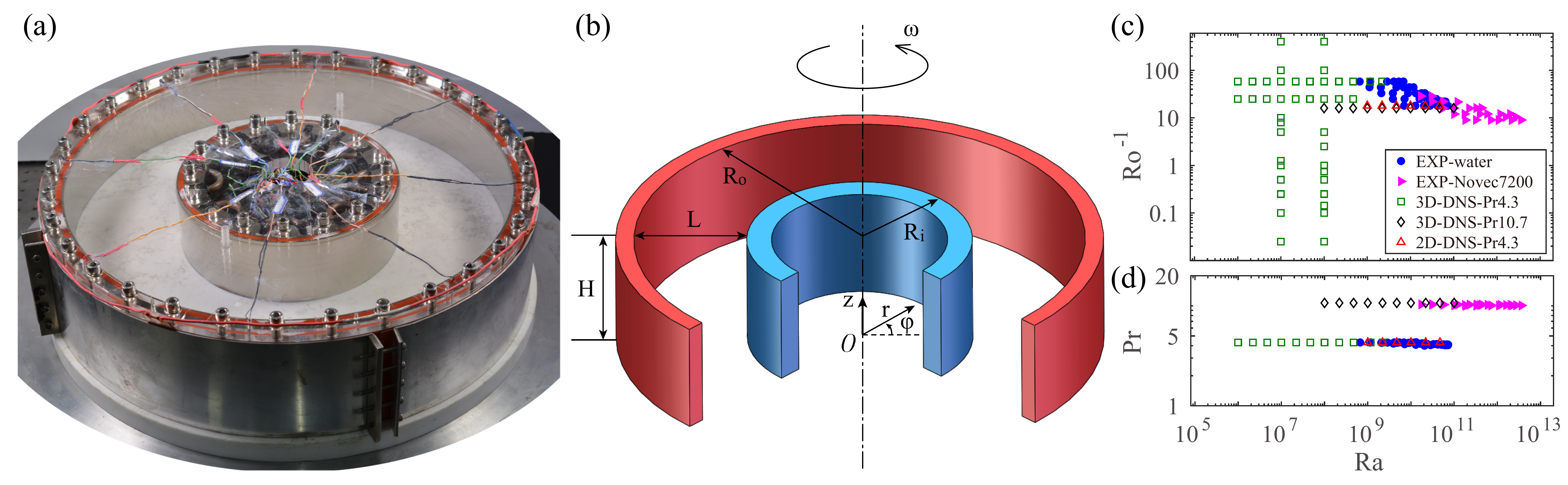}}
	\caption{
		Experimental configuration and explored parameter space. (a) A photograph of the annular convection cell. (b) Schematical diagram of the system, which defines the coordinate frame and geometric parameters. (c) Phase diagrams of $\text{Ro}^{ - 1}$ and $\text{Ra}$, where solid symbols and open symbols denote the experimental data and numerical data, respectively. We have also performed numerical simulations with negligible Coriolis force at $\text{Ro}^{-1}=10^{-5}$ (not shown here). (d) Phase diagrams but in the  $\text{Pr}-\text{Ra}$ plane.
	}
	\label{Fig1}
\end{figure*}

In order to study the ultimate thermal turbulence, we propose to use a supergravitational system (Annular Centrifugal RBC, ACRBC) (sketched in figures \ref{Fig1} (a,b)), which is a cylindrical annulus with cooled inner and heated outer walls under a rapid solid-body rotation \cite{JiangSciAdv,rouhi2021}. { The direction of centrifugal force (effective gravity) is the same as the temperature gradient, which is consistent with classical RBC.}  {The details about the experiments and direct numerical simulations can be found in Method section below.  } 

The flow dynamics in ACRBC are determined by the following control parameters, namely the Rayleigh number
\begin{equation}
	\text{Ra} = \frac{1}{2}\omega ^2({R_o} + {R_i})\beta \Delta {L^3}/(\nu \kappa ),
\end{equation}
the inverse Rossby number
\begin{equation}
	\text{Ro}^{-1}=2{(\beta \Delta ({R_o} + {R_i})/(2L))^{ - 1/2}}, 
\end{equation}
the Prandtl number $\text{Pr}  = \nu /\kappa $, and radius and aspect ratios $\eta=R_i/R_o$, $\Gamma _ \bot = H/L$ and $\Gamma _\parallel= 2\pi r/L$. Here $\beta$, $\Delta$, $\nu$ and $\kappa$ are the thermal expansion coefficient, the temperature difference between hot and cold walls, the kinematic viscosity and the thermal diffusivity of the working fluid, respectively. The key response parameter is Nusselt number $Nu = J/J_{con} = -JR_o\ln \eta/(\alpha \Delta )$, where $J$, ${{J_{con}}}$ and $\alpha$ denote the total heat flux, the heat flux through pure thermal conduction and thermal conductivity, respectively. 

According to the definition of Rayleigh number, we can effectively push Ra to higher values through increasing rotation rate $\omega$ of the system and increasing $\beta/(\nu\kappa)$ of the working fluid. {The range of rotation rate of the system is from 158rpm to 705rpm, which corresponds to an effective gravity [5g,100g]. The Earth's gravity does not play an important role in the current parameter regime (see \cite{JiangSciAdv} for detailed discussion)} Next to the classical degassed water at around ${40}{\celsius}$, we use Novec 7200 (3M Inc. Engineered Fluid) at around ${25}{\celsius}$ as the working fluid, which has roughly 14.4 times of  $\beta/(\nu\kappa)$ as compared with water. The properties of the Novec 7200 engineered fluid are listed in Suppl. Mat. for reference.

The explored parameter space of experimental and numerical studies is shown in figure \ref{Fig1} (c). We have performed 72 experiments and 67 numerical simulations in total. The details about the experimental and numerical cases are documented in the Suppl. Mat. Combining experiments and simulations, the range of Rayleigh number explored here extends almost six and a half decades, i.e. from ${10^6}$ to $3.7 \times {10^{12}}$. In experiments, the Earth's gravity, lids and inhomogeneity of centrifugal force have been already studied in \cite{JiangSciAdv}, which shows that their effects on Nu are negligible.

\section{Results and discussions}

\begin{figure*}[htp]
	\centerline{\includegraphics[width=0.99\linewidth]{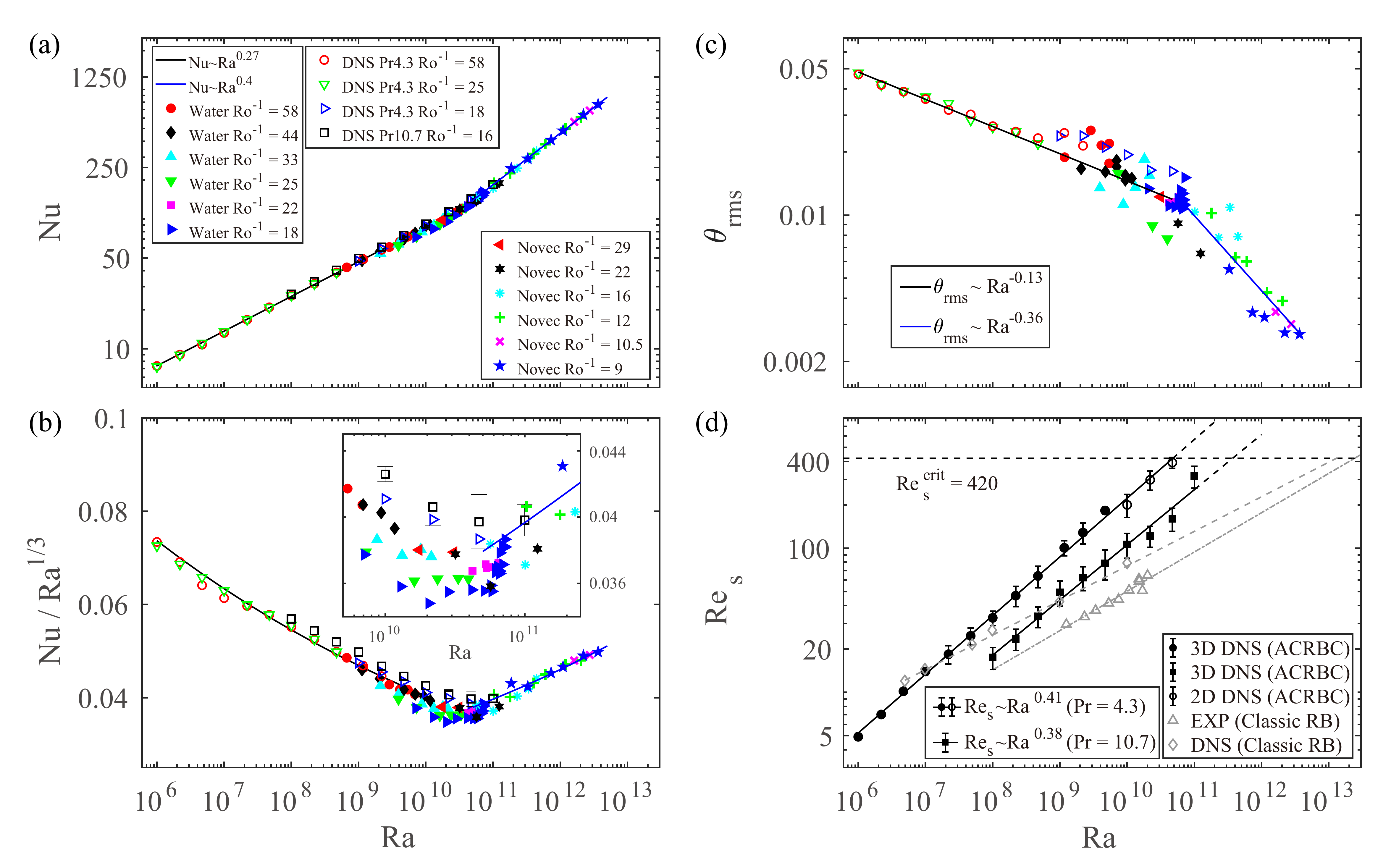}}
	\caption{
		Global heat transport, temperature fluctuation and shear Reynolds number.  (a) Nusselt number Nu as a function of Rayleigh number Ra; (b) the compensated plots of $\text{Nu}/\text{Ra}^{1/3}$ versus Ra; (c) the normalized r.m.s. of temperature fluctuation ${\theta _{rms}}$ in the bulk region versus Ra; (d) the shear Reynolds number $\text{Re}_s$ as a function of Ra. In panel (a), the solid lines are the best fittings of the experimental and DNS data in ACRBC at two different regimes. To eliminate the weak influence of Pr difference, Nusselt numbers  are corrected with $\text{Nu}/{({\Pr}/4.3)^{ - 0.03}}$ \cite{Ahlers2001,grossmann2001thermal,Xia2002PRL}. The inset shows the enlarged part near the transition Ra.
		In panel (d), the extrapolations of the fitting of DNS results for water ($\text{Ro}^{-1}$=18 and 58) and Novec ($\text{Ro}^{-1}$=16) indicate that the BL becomes turbulent (${{\mathop{\rm Re}\nolimits} _s} = 420$, dashed line) at $\text{Ra} \simeq {10^{11}}$.  Shear Reynolds numbers from experiments \cite{Sun2008} and from DNS \citep{sch14} in classical RBC system are also plotted for comparison.} 
	\label{Fig2}
\end{figure*}

Figure \ref{Fig2} (a) shows the obtained Nu as a function of Ra from experiments and numerical simulations. For most experiments, the measurement lasts at least 4 hours after the system has reached statistically stationary state (for the detailed measuring procedure, see Suppl. Mat.). Since two kinds of working fluids are used in the experiments, the difference in Prandtl number should be taken into account. As previous study suggested that Nu has a weak dependence on Pr in this Pr range $[4,10.7]$ and the scaling law can be written as $\text{Nu} \sim \Pr^{-0.03}$ \cite{Ahlers2001,grossmann2001thermal,Xia2002PRL}, therefore all Nu data, including the data at $\text{Pr} \approx 10.4$ (Novec 7200) and the data at $\text{Pr} \approx 4.3$ (water), are corrected with $\text{Nu}/{({\Pr}/{4.3})^{ - 0.03}}$ to coincide with the data for water at ${40}{\celsius}$. It is evident that the experiments for Novec 7200 and water and numerical simulations are all in an excellent agreement. In the range of $\text{Ro}^{-1}$ from 9 to 58, the data sets show a consistent dependence of Nu on Ra. 
To better demonstrate the local scaling exponent, figure \ref{Fig2} (b) shows the same plot as figure \ref{Fig2} (a) but in a compensated way. In ACRBC, an effective scaling of $\text{Nu} \propto \text{Ra}^{0.27}$ is observed when $\text{Ra}<10^{10}$, and the scaling exponent is close to the value found in two-dimensional (2D) RBC \cite{van_der_poel2013,Zhu2018PRL} where the viscous BLs are laminar.
Note, in the current system, the flow has a quasi-2D structure at the current $\text{Ro}$ range ($\text{Ro}^{-1} \geqslant 9$) for Nusselt number measurements \cite{JiangSciAdv}. 
{The findings of classical state regime in ACRBC give independent support for the previous results on RBC \cite{Roche2010,Urban2011} and Taylor-Couette turbulence \cite{ost14pd,Froitzheim2019}.}

Once the Ra increases beyond $10^{10}$, the system enters a transition regime with local effective scaling exponent $\gamma$ of $\text{Nu}\propto \text{Ra}^{\gamma}$ increasing from $\gamma=0.27$ to $\gamma>1/3$ as evident from the arc bottom in the compensated plot. Surprisingly, following the transition regime, there is a steep scaling regime with a local scaling exponent $\gamma=0.40\pm0.01$, {which spans more than one decade from $\text{Ra} \sim 10^{11}$ to $\text{Ra}=3.7\times10^{12}$.} This steep scaling exponent is consistent with the prediction for $\text{Nu}(\text{Ra})$ in the ultimate regime with logarithmic corrections $\text{Nu} \propto \text{Ra}^{1/2}{(\log \text{Ra})^{ - 3/2}}$ \cite{Kraichnan1962}. 

What is the reason for the enhanced Nu scaling? Could this be due to the transition of the flow structure from a two-dimensional flow state to a three-dimensional state, which results in a locally higher Nu?
As shown in figure 15 in Suppl. Mat., the Nu for the steep scaling regime is much higher than the Nu extrapolated for the case at $\text{Ro}^{-1}=10^{-5}$, in which the flow has a three-dimensional state, indicating that the enhancement of local scaling exponent should not be attributed to the change of flow state from two-dimensional to three-dimensional. It suggests the physical reason might be that the transition Ra to the ultimate regime in ACRBC is lower than that in the traditional RBC (to be discussed below).

The normalized root-mean-square temperature fluctuation ${\theta _{rms}}$ in the bulk region as a function of Ra is illustrated in figure \ref{Fig2} (c). It is evident that there is a transition in the variation trend of ${\theta _{rms}}$ vs. Ra at exactly the critical Rayleigh number $\text{Ra}^{*}\simeq10^{11}$ where the Nu scaling changes. The obtained scaling ${\theta _{rms}}\sim\text{Ra}^{ - 0.13 \pm 0.02}$ for $\text{Ra} < \text{Ra}^{*}$ is consistent with previous studies for the classical RBC \cite{Castaing1989,Niemela2000,Yang2020}, while it gives a notably steeper scaling with ${\theta _{rms}}\sim\text{Ra}^{ - 0.36 \pm 0.05}$ when $\text{Ra}>\text{Ra}^{*}$. The bulk temperature fluctuation is suggested to be dominated by contributions of detached plumes from the BLs \cite{Castaing1989}. This change of the scaling exponent of  ${\theta _{rms}}$ versus Ra in the {Kraichnan-like ultimate regime} indicates the significant change in the properties of the BL where the plumes are emitted.

The next key question is that why the transition Rayleigh number ($\text{Ra}^* \sim 10^{11}$) is three orders of magnitude lower than that in the traditional RB system ($\text{Ra}^* \sim 10^{14}$)?

{
	The key ingredient of the transition from classical regime to ultimate regime is the change of BL properties, i.e. the transition from laminar BL to turbulent one \cite{Kraichnan1962,grossmann2000,gl2011}. The properties of BL critically depend on the shear Reynolds number.
	A typical value for the onset of turbulence is $\text{Re}_s^{\text{crit}}\approx 420$ \cite{Landau1987}. }	
We now analyze $\text{Re}_s$ as a function of Ra in the current system and compare it to that in the classical RBC system. It is difficult to directly measure the shear Reynolds number from the experiments due to the rapid rotation and very small scale of the BL. Fortunately, we can evaluate $\text{Re}_s$ using the DNS data. 
Figure \ref{Fig2} (d) shows the calculated shear Reynolds number, ${{\mathop{\rm Re}\nolimits} _s} = U{\lambda _u}/\nu$, based on the DNS results, here $U$ is the maximum of temporally and spatially averaged azimuthal velocity, ${\lambda _u}$ the viscous BL thickness estimated with the commonly used `slope method' \cite{zhou10}. 
As shown in figure \ref{Fig2} (d), the shear Reynolds number monotonously increases with Ra with an effective scaling around {$\text{Re}_s \sim \text{Ra}^{0.41\pm0.01}$} for Pr = 4.3 and {$\text{Re}_s \sim \text{Ra}^{0.38\pm0.01}$} for Pr = 10.7. These scaling exponents are much larger in the current system than that in the classical RBC. {The magnitude of $\text{Re}_s$ is also much larger than that in the classical RBC.
	It already arrives at the critical shear Reynolds number of 420 \cite{Landau1987} at $\text{Ra} \simeq {10^{11}}$, whereas the shear Reynolds number at the same Ra in the classical system is around 80 \cite{Sun2008,sch14}. 
	
	
\begin{figure}
	\centerline{\includegraphics[width=0.99\linewidth]{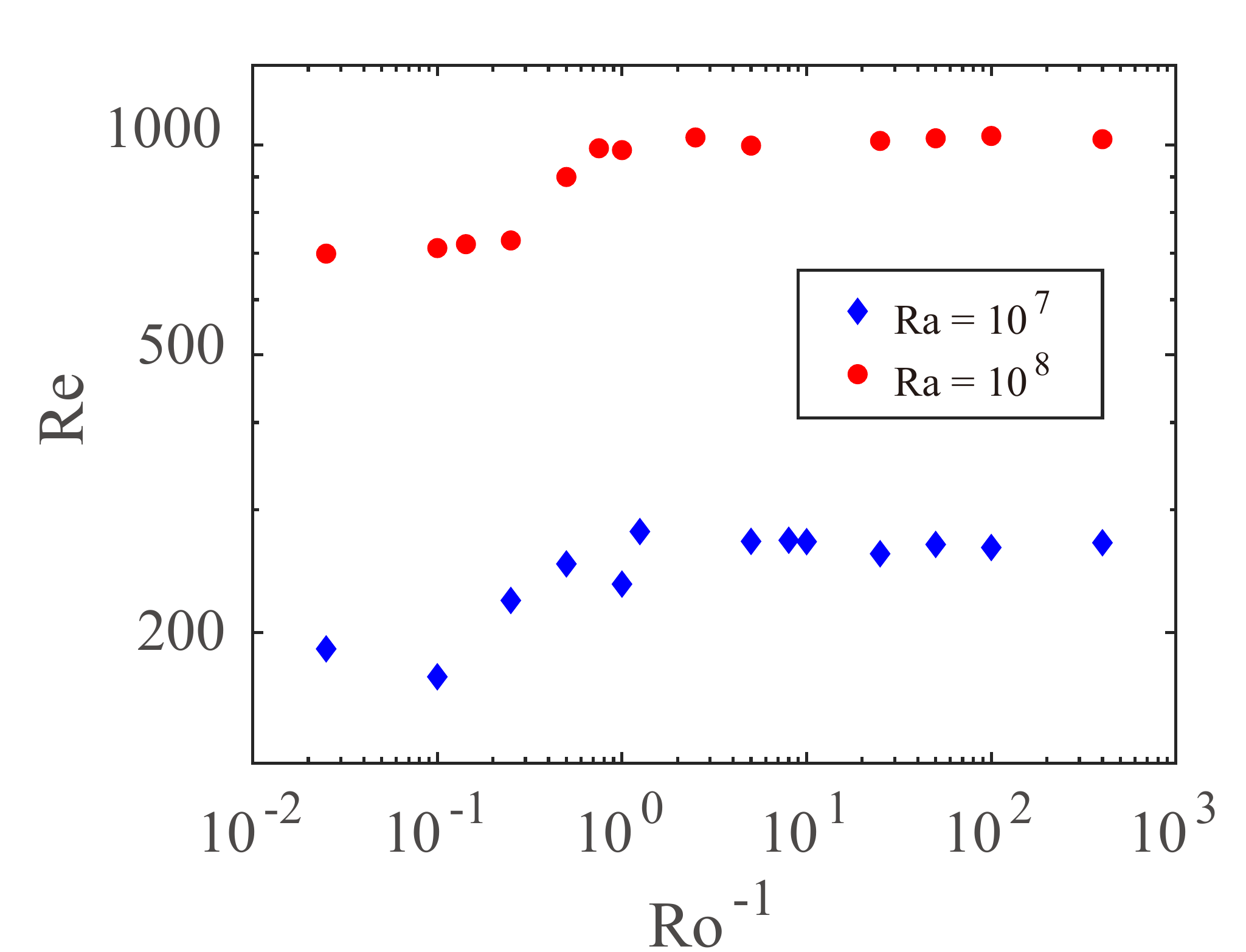}}
	\caption{
		{Reynolds number $\text{Re}=U_{rms}L/\nu$ as a function of $\text{Ro}^{-1}$ at $\text{Ra}=10^7$ and $10^8$ for $\text{Pr}=4.3$.}
	}
	\label{Fig3}
\end{figure}

{To understand the large magnitude and steep scaling of the shear Reynolds number in rapidly rotating ACRBC, we show in figure \ref{Fig3} how the Reynolds number $\text{Re}=U_{rms}L/\nu$ is influenced by rapid rotation. The dependences of $\text{Re}$ with $\text{Ro}^{-1}$ at two specified Rayleigh numbers, $\text{Ra}=10^7$ and $10^8$, demonstrate a consistent behavior that $\text{Re}$ first increases with $\text{Ro}^{-1}$ and then saturates, showing that under rapid rotation the flow strength is enhanced, which will yield a larger shear Reynolds number $\text{Re}_s$ and promote the transition of laminar viscous BL to be turbulent. 
	Regarding the scaling of $\text{Re}_s$ with $\text{Ra}$, we consider the force balance of the BL flow based on the momentum transport equation in the large scale circulation plane.
	For traditional RBC, the balance of the inertial term $\vec{u}\cdot\nabla\vec{u}$ and the viscous term $\nu\nabla^2\vec{u}$ gives the classical scaling of the viscous BL, ${\lambda _u}/L\sim\text{Re}^{-1/2}$ and $\text{Re}_s\sim\text{Re}^{1/2}$. Whereas in ACRBC, when rotation is rapid enough such that the Coriolis force $2\vec{\omega}\times\vec{u}$ plays an important role in the BL flow, one would expect a balance between  the Coriolis force and the viscous force, which results in a new scaling behavior, ${\lambda _u}/L\sim\text{Re}^{-1/3}$ and $\text{Re}_s\sim\text{Re}^{2/3}$. 
	Together with the effective scaling law $\text{Re}\sim\text{Ra}^{0.55}$, the new scaling gives rise to the scaling law $\text{Re}_s\sim\text{Ra}^{0.37}$, close to the observation in figure \ref{Fig2} (d). {The detailed derivation of $\text{Re}_s\sim\text{Ra}$ scaling can be referred to the Suppl. Mat.} Thus, the dominance of Coriolis force yields the steep scaling of $\text{Re}_s$ with $\text{Ra}$ and thus results in the early transition of ACRBC to the ultimate regime at $\text{Ra}^*\sim10^{11}$.

\begin{figure}
	\centerline{\includegraphics[width=0.99\linewidth]{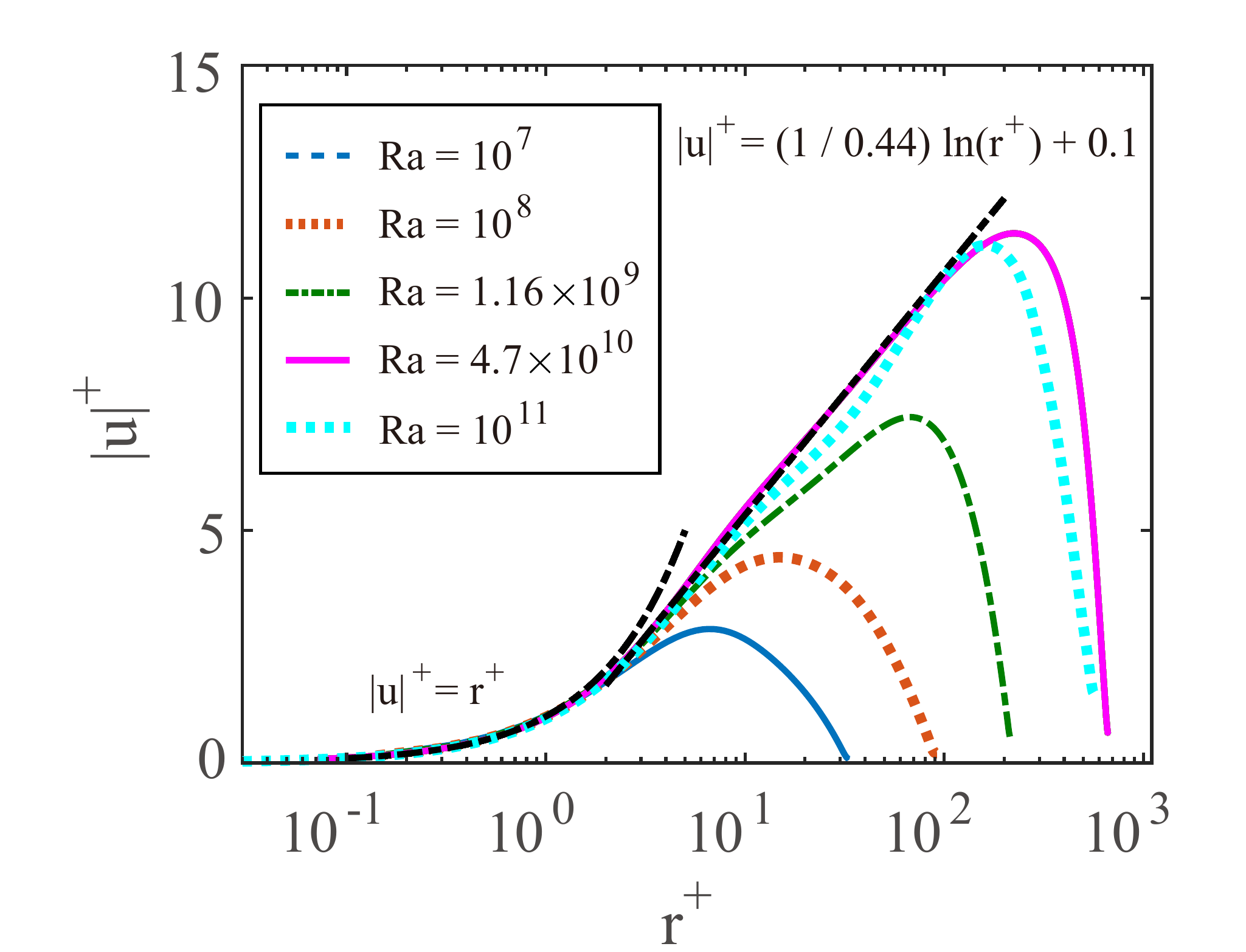}}
	\caption{
		Mean boundary layer velocity profiles in wall units at different Ra, Ra = $10^7-4.7\times10^{10}$ ($\text{Ro}^{-1}=$ 18, 58 and Pr = 4.3) and Ra = $10^{11}$ ($\text{Ro}^{-1}=$ 16 and Pr =10.7), $u^{+}$ for velocity and $r^{+}$ for wall distance. The dash-doted lines show the viscous sublayer behavior $|u{|^ + } = {r^ + }$ and log-layer behavior {$|u{|^ + } = (1/\kappa)\ln ({r^ + }) + B$ with $\kappa=0.44$ and $B=0.10$}.  
	}
	\label{Fig4}
\end{figure}

Figure \ref{Fig4} shows the temporally and spatially averaged azimuthal velocity profiles in wall units for different Ra from DNS. 
In wall units, the mean azimuthal velocity $|u{|^ + }$ is normalized by the friction velocity ${u_\tau } = \sqrt {\nu {\partial _r}\left\langle u \right\rangle {|_{{R_i},{R_o}}}}$, and radial position ${r^ + }$ is normalized by the viscous length scale ${\delta _\nu } = \nu /{u_\tau }$ \cite{Pope2000}. The profiles of viscous sublayer ($|u{|^ + } = {r^ + }$) and Prandtl-von K\'{a}rm\'{a}n type BL {($|u{|^ + } = (1/\kappa)\ln ({r^ + }) + B$} \cite{Yaglom1979} are also plotted for comparison. We show that when Ra is small, the profile behaves as the laminar Prandtl-Blasius type. With increasing Ra, the BL profile progressively approaches towards a Prandtl-von K\'{a}rm\'{a}n (logarithmic) type.
At {$\text{Ra} = 4.7 \times {10^{10}}$ and ${10^{11}}$}, {a logarithmic range is notable over one decade of ${r^ + }$, which is an essential indication of the emergence of turbulent BL. The inverse slope $\kappa=0.44$ of the obtained logarithmic law is quite close to the typical von K\'{a}rm\'{a}n constant $\kappa=0.41$ despite the different flow configurations. The parameter $B=0.10$ is different from the classical Prandtl-von K\'{a}rm\'{a}n value for canonical turbulent BLs over smooth walls, which may attribute to the finite values of $\text{Ra}$ reached in the DNS and the complicated interactions between velocity shear, unstable thermal stratification, the Coriolis force, and curvature effect in ACRBC. Nevertheless, the fact that the velocity profile shows the typical characteristics of a turbulent BL illustrates the onset of turbulence in the BL flow.}
Unfortunately, higher Ra simulation has not been achieved in the present study. Above analysis gives an indication that the strong shear and Coriolis effects induced by rapid rotation promote the BL transition to the turbulent Prandtl-von K\'arm\'an type \cite{rouhi2021}.
\section{Conclusions}
By means of dramatically increased driving force and strong shear induced by rapid rotation, the transition to the {Kraichnan-like ultimate regime} of thermal convection is observed in an annular centrifugal Rayleigh-B\'{e}nard convection system.
We have performed extensive experiments and numerical simulations to study the heat transport and flow dynamics in ACRBC from classical regime to ultimate regime. For $\rm{Ra} \lesssim {10^{10}}$, $\text{Nu}(\text{Ra})$ is consistent with classical RBC as expected. For $\text{Ra} \gtrsim {10^{11}}$, the measured local effective Nu scaling exponent $\gamma$ increases to 0.40, {spanning more than one decade of Ra range, which testifies the possible transition to ultimate regime.} As a response to the transition to ultimate regime, the dependence of temperature fluctuations on Ra demonstrates different scaling behaviors before $\text{Ra}^{*}$ and beyond $\text{Ra}^{*}$. The steep  Ra-dependence and elevated amplitude of shear Reynolds number lead to a smaller transition Rayleigh number $\text{Ra}^{*}$ at which the shear Reynolds number crosses $\text{Re}_s=420$.  Approaching the transition $\text{Ra}^{*}$, the mean velocity profile has a log layer spanning over one decade of $r^+$ in wall units.  
In view of these above evidences, it could be concluded that transition to the {Kraichnan-like ultimate state} of thermal convection has been realized using supergravitational thermal convection system. {Of course, more studies are needed to further verify this transition to ultimate regime of thermal turbulence.}
\section{Methods}

\subsection{Experiment}

Since the experimental setup has already been described detailedly in \cite{JiangSciAdv}, here we just provide some main features associated with our study. As sketched in figure \ref{Fig1} (a), the experiments are conducted in a cylindrical annulus with solid-body rotation around its vertical axis. The cylindrical annulus consists of two concentric cylinders, which are machined from a piece of copper to keep dynamically balanced. To prevent oxidation, the surfaces of cylinders are electroplated with a thin layer of nickel. As shown in figure \ref{Fig1} (b), the inner radius of the outer hot cylinder is ${R_o} = {240}{\milli\metre}$, and the outer radius of the inner cold cylinder is  ${R_i} = {120}{\milli\metre}$, resulting in a gap of width $L=R_o-R_i={120}{\milli\metre}$ and a radius ratio of $\eta  = {R_i}/{R_o} = 0.5$. The cylindrical annulus with a height of $H={120}{\milli\metre}$ is sandwiched by a plexiglass lid and a teflon bottom, hence resulting in an aspect ratio in the large scale circulation plane $6 \lesssim {\Gamma _\parallel }= 2\pi r/L \lesssim 12$ and in the plane perpendicular to the large scale circulation $\Gamma _ \bot = H/L=1$. These cylinders and end plates are fixed together and leveled on a rotary table, which can rotate at a modulated speed ($\omega$). 
{Here, we note that since the end surfaces of ACRBC are flat and the boundary condition on the cylindrical surfaces is no-slip, zonal flow is very weak in the present study \cite{busse_1974,busse_1985,busse_1986a,busse_1986b,busse_1994}. Thus, it is considered that the rotating convection and non-rotating convection are roughly similar in terms of flow dynamics and transport mechanism.}  
A high-precision circulating bath (PolyScience Inc., AP45R-20-A12E) pumps cold circulating water to maintain a constant temperature of inner cylinder, and four silicone rubber film heaters supplied by a DC power (Ametek Inc., XG 300-5) are connected in series and attached to the outside of the outer cylinder to provide a constant and uniform heating. For high-precision thermal experiments, special measures are taken to minimize the heat exchange between the system and the surroundings. {Note that the temperature difference in our experiments is small (from 2K to 20K) and the Oberbeck-Boussinesq conditions are well satisfied \cite{ahlers_brown_araujo_funfschilling_grossmann_lohse_2006}.}

\subsection{Governing equations and numerical simulations}

The parameter definitions are shown in figure \ref{Fig1}(b). {We employ the Boussinesq equations to describe the fluid flow, which are commonly adopted in the theoretical, numerical and experimental studies of thermal convection \cite{Review1}.} The dimensionless Boussinesq equations in the rotating reference frame \cite{lop13} are expressed as:

\begin{small}
	\begin{align}
		\nabla \cdot {\vec u} &= 0, \\
		\frac{\partial \theta}{\partial t} +{\vec u}\cdot \nabla \theta &= \frac{1}{\sqrt{\text{Ra}\text{Pr}}}\nabla^2 \theta, \\
		\frac{\partial \vec{u}}{\partial t} +{\vec u}\cdot \nabla {\vec u} &= -\nabla p+\text{Ro}^{-1} \hat{\bm{\omega}} \times {\vec u} \nonumber \\ 
		&\phantom{=} +\sqrt{\frac{\text{Pr}}{\text{Ra}}}\nabla^2 {\vec u} -\theta \frac{2(1-\eta)}{(1+\eta)}\vec r,
	\end{align}
\end{small}
where $\vec u$ is the normalized velocity vector using the free fall velocity of the system $\sqrt{\omega ^2\frac{R_o+R_i}{2}\beta \Delta L}$, $\bm{\hat{\omega}}$ the unit vector in the angular velocity direction, $\theta$ the temperature that is normalized by the temperature difference $\Delta = T_h - T_c$, and $t$ the dimensionless time that is normalized with $\sqrt{L/(\omega ^2\frac{R_o+R_i}{2}\beta\Delta)}$.
Here, $\omega$ and $\Delta$ are the rotational speed, and the temperature difference of the two cylinders with the hot wall temperature $T_h$ and the cold wall temperature $T_c$, respectively.

Direct numerical simulations (DNS) are carried out in the cylindrical coordinates by using the redeveloped second-order finite-difference AFiD code \cite{VERZICCO1996402,VANDERPOEL201510,zhu18a}, which has been extensively validated in prior work on turbulent flows \citep{zhu18a,zhu18b,JiangSciAdv}. The geometry of the calculated domain is chosen to be the same as used in experiments except for the high-Ra cases where a computational domain with reduced height is adopted. It is appropriate due to the quasi-geostrophy of the flow field under rapid rotation. The no-slip boundary conditions for the velocity and constant temperature boundary conditions are adopted at the inner and outer cylinders. Periodical conditions are prescribed for the velocity and temperature in the axial direction. Adequate grid resolutions and statistical convergence are ensured for all cases. At the highest $\text{Ra} = 10^{11}$, $18432\times1536\times48$ grid points are used and most simulations were run for at least additional 100 free-fall time units after the system has reached statistically stationary state. See Suppl. Mat. for more numerical details of the simulations.


\bigskip

\begin{acknowledgments}
\textbf{Acknowledgments:} We thank Guenter Ahlers, Eberhard Bodenschatz, Detlef Lohse, Roberto Verzicco, Ke-Qing Xia, Yantao Yang, Quan Zhou and Xiaojue Zhu for insightful discussions over the years, and thank Gert-Wim Bruggert and Sander Huisman for the technical assistance with the setup.
This work is financially supported by the Natural Science Foundation of China under Grant No. 11988102, 91852202, 11861131005, and Tsinghua University Initiative Scientific Research Program (20193080058).

\bigskip

\textbf{Competing interests}: The authors declare that they have no competing interests.

\bigskip

\textbf{Data and materials availability:}: All data needed to evaluate the conclusions in the paper are present in
the paper and/or the Supplementary Materials. Additional data related to this paper may
be requested from the authors.

\end{acknowledgments}

\end{document}